# Large-scale dynamic assembly of metal nanostructures in plasmofluidic field

Partha Pratim Patra, Rohit Chikkaraddy[¶], Sreeja Thampi, Ravi P. N.Tripathi, G.V. Pavan Kumar*

Photonics & Optical Nanoscopy Laboratory, h-cross,

Indian Institute of Science Education and Research, Pune – 411008, INDIA

*corresponding author's e-mail: pavan@iiserpune.ac.in

**Abstract**

We discuss two aspects of plasmofluidic assembly of plasmonic nanostructures at metal-fluid interface. First, we experimentally show how triple and four spots evanescent-wave excitation can lead to unconventional assembly of plasmonic nanoparticles at metal-fluid interface. We observed that the pattern of assembly was mainly governed by the plasmon interference pattern at the metal-fluid interface, and further lead to interesting dynamic effects within the assembly. The interference patterns were corroborated by 3D finite-difference time-domain simulations. Secondly, we show how anisotropic geometry, such as Ag nanowires, can be assembled and aligned in unstructured and structured plasmofluidic fields. We found that by structuring the metal-film, Ag nanowires can be aligned at metal-fluid interface with a single evanescent-wave excitation, thus highlighting a prospect for assembling plasmonic circuits in a fluid. An interesting aspect of our method is that we obtain the assembly at locations away from the excitation points, thus leading to remote assembly of nanostructures. The results discussed herein may have implications in realizing a platform for reconfigurable plasmonic metamaterials, and a test-bed to understand the effect of plasmon interference on assembly of nanostructures in fluids.

¶ Current address: University of Cambridge

**Introduction**

Manipulation and reversible assembly of large number of nanoparticles in a fluid is of relevance in various aspects of microfluidic science and technology.[1-7] The assembly of nanoparticles, especially of noble metals, has also derived attention in self-assembled plasmonic metamaterials[8-14] that can control various parameters of light at sub-wavelength scales. An inherent challenge to overcome while handling nanoparticles in liquid is the Brownian motion. Various possible methods have been developed to overcome this hindrance, of which optical trapping has attracted significant attention.[15-21] Although laser-based conventional optical trapping has been very successful in trapping and manipulating sub-micron scale objects, it still needs powerful lasers to create gradient and scattering forces at a tightly focused location[22-24]. This requirement of large optical power is detrimental to trapping of soft nanomaterials, as they can severely damage the structure of the trapped object. An alternative method of trapping sub-micron scale objects with low power laser excitation is to utilize the evanescent field of surface plasmons at metal-dielectric interface[25-31]. Surface plasmons - the collective oscillation of free electrons and light at metal-dielectric interface - can localize optical fields to subwavelength scale and propagate optical signals beyond diffraction limit[32-37]. This capability to localize and propagate light can be further harnessed to trap and manipulate large assemblies of nanoparticles, including resonant plasmonic nanoparticles,[29, 38-40] at metal-fluid interface. Such trapping methods which use plasmonic field in fluidic environments are generally called plasmofluidic traps,[6] and can be harnessed to trap and manipulate nano and microscale objects in the presence of Brownian motion.

Recently, our group has developed a method[41] to trap and manipulate plasmonic nanoparticles assembly at an unstructured metal-fluid interface. We showed that, by using single evanescent optical excitation, our trapping mechanism could not only manipulate nanostructures in a fluid, but also detect single-molecule Raman scattering signatures in the created trap. The same method was also shown to work with dual evanescent wave excitation, which was harnessed to show interaction between dynamic assemblies of nanoparticles. Our method, in an essence, has expanded the capability of dynamic trapping and single-molecule assisted probing of nanoparticle assemblies using surface plasmons. Furthermore, there are still many interesting questions regarding our method that are to be addressed such as: a) What will be the effect of multiple evanescent plasmon excitation on dynamic assembly of plasmonic nanoparticles in fluids? Answering this question will lead to understanding of how nanoparticle assemblies interact with each other at metal-fluid interface, especially when the interaction is mediated by the near-field interference of surface plasmons. b) Can we extrapolate our method to trap anisotropic geometries such as plasmonic nanowires? The answer to this question will open a new possibility of manipulating and arranging nanowires in liquids, which can be further harnessed for nano-optical and nano-optoelectronic circuitry in liquids.

In this paper we present experimental results that address the above-mentioned questions for some special cases. We show a) how three and four evanescent-optical excitations can lead to interesting assemblies of plasmonic nanoparticles at metal-fluid interface; and b) how plasmonic Ag nanowires can be trapped and aligned at unstructured and structured metal-fluid interface.

**Experimental section**

**Preparation of unstructured and structured plasmonic thin film:**

The unstructured gold thin films were prepared by Direct current (DC) plasma sputtering technique, already discussed in our previous work. The thickness of the film was $(50\pm5)$ nm.

The structured gold thin films were produced by photolithography and followed by Au deposition through plasma sputtering. First, a glass substrate was thoroughly cleaned and heated at 80-100°C. Then, it was coated with the photo-resist at 4000 rpm for 3mins. The photo-resist coated glass substrate was baked at 85°C. Thereafter, the patterns were written using 405 nm laser line. After developing the glass substrate, 50 nm Au was deposited on to it by DC sputtering. Finally this was washed with acetone to remove the remaining photoresist to make ready for the assembly experiment. The geometry of the structured film was like periodic strips; each strip was of length-50 µm, width-2 µm, thickness-$50(\pm5)$ nm and the gap between two consecutive strips was 2 µm.

**Synthesis of Silver nanoparticles and nanowires:**

In this work we used aqueous phase silver colloidal nanoparticles, synthesized by citrated reduced method.[42] 45 mg $AgNO_3$ was dissolved in 250 ml water and heated. While boiling 5mL of 1% sodium citrate aqueous solution was added and kept boiling for 1 hr.

The silver nanowires were synthesized by polyol process.[32, 43] 3mL of each, 0.1M $AgNO_3$ solution and 0.6M polyvinylpyrrolidone (PVP, MW, 55000) solution in ethylene glycol (EG) were mixed together in room temperature. This mixture is then injected drop wise to 5mL preheated EG at 160°C and kept it for 1 hr. The nanowires then washed in acetone and water. The final Ag nanowire suspension was made in water.

**FDTD simulations:**

The simulation set up consisted of a 50 nm thin gold film placed on a glass slab and the medium above the metal film was modeled with refractive index of 1.34 (water). A vertical dipole (polarization perpendicular to the plane of metal film) of 532 nm was placed at the interface of metal and glass for the excitation of SPPs on metal film. Two-dimensional (2D) frequency domain power monitor was set at the metal-water interface to record the near-field image. Perfectly matched layer (PML) boundary conditions were applied in the far-field domain to avoid reflections from the boundaries. For multiple excitations, equivalent numbers of vertical dipoles were placed at the appropriate (xy-plane) position at the metal-glass interface.

To model the structured film (metallic strip) experiment, simulations were done for a vertical dipole excitation at the center of a single strip (dimensions matching with experimental parameters) with periodic boundary condition in x-direction and PML on the other two axes. The difference between the excitation methods between the experiments and simulation was justified since only the z-component of electric-field (p-polarization) contributes to the SPP excitation.

**Result and Discussion**

**Plasmofluidic assembly of metal-nanoparticles:**

This work is a follow-up of our previous work[41] on single molecule surface-enhanced Raman scattering from plasmofluidic assembly of nanoparticles in evanescent excitation geometry. Before the discussion on new results, the schematic illustration is shown in Fig. 1 as a preamble of this work.

**Multiple-spot-excited plasmofluidic assembly of Ag nanoparticles:**

We first addressed the issue of how to excite multiple evanescent-waves at metal-fluid interface to assemble metal nanoparticles. In order to do so we chose to excite multiple surface plasmon waves on metal film in Kretschmann geometry (Fig. 2a). Choice of nanoparticles, metal film and wavelength of excitation are crucial, so we utilize the optimized parameters from our previous study.[41]

The plasmon excitation set-up (Fig. 2a) consists of 532-nm continuous-wave frequency-doubled Nd:YAG laser (maximum power up to 200 mW) coupled with a Dove-prism (N-BK7, Refractive index, 1.519). A glass coverslip coated with gold thin film (50±5 nm) was used as the substrate. The p-polarized light was used at an incidence angle of 72.7° for the assembly process. The multiple beams with equal power were created by using 50-50 beam splitters (BS), polarizing beam-splitters (PBS) and half-wave plates ($\lambda/2$). Four weakly focusing lenses of same focal length were used to focus the light. . In such geometry the power density at the excitation spot arises as approximately 100W/cm$^2$ which is very low compared to the conventional objective lens based excitation ($\sim 10^6$ W/cm$^2$). To capture the real time assembly process, objective lenses (OBL) with magnification $\times 4$ and $\times 60$ were used. To capture bright field images a white light source was combined with the imaging system and a high pass filter (HPF) was introduced before the CCD camera to block the excitation beam. In Fig. 2b the dark-field image of the triple-spot-excited gold thin film is shown.

Onto this film 200 μL Ag colloidal solution was drop-casted and the assembly process was recorded in real time (movie-1 in ESI). Fig. 2c is a bright-field snap shot of the assembly process taken after 12 minute of excitation. We observed that the nanoparticles were assembled at the excitation spots as well as at the orthocenter where there was no direct optical-excitation. In our earlier study,[41] we elaborately explained the role of hydrodynamic and optical forces in the assembly of nanoparticles at the single excitation spot. Here, convoluted effect of multiple excitations resulted in the formation of nanoparticles-assembly at the orthocenter and excitation points.

To understand the assembly pattern, we mapped the electric fields at metal-fluid interface through full-wave 3-D FDTD simulations. The near-field |E| map is shown in Fig. 2d. We see that the electromagnetic fields from the propagating plasmons of the metal film interact with each other. The collective interference of SPP waves from three excitation points results in a unique pattern of nearfield at the metal-fluid interface, which in turn governs both optical and hydrodynamic forces on the nanoparticles. The schematic of the interaction of the forces is shown in Fig. 2e. At the orthocenter, two predominant forces get involved but in opposite directions: one is gradient force due to laser excitation and other is the convective force due to heating by the same.[44-46] Thus, the assembly at orthocenter is due to the balance between two counter steering forces in the plasmofluidic environment. To explain the nanoparticle chain formation, we primarily attribute the SPP interference (Fig. 2d) at the metal -fluid interface, though in such plasmofluidic systems all the effects are codependent and thus complex in nature.

**Kinetics of triple-spot-excited plasmofluidic assembly of Ag nanoparticles:**

To obtain further insight into the assembly process, we analyzed the variation of nanoparticles density and position with time. In Fig. 3a the growth kinetics of the nanoparticles assembly at the excitation points is shown. The growth plot fits well with a sigmoidal curve. It took only few minutes (4-5 min) for the nanoparticles to assemble and after a particular time the assembly process seems to be saturated. This type of assembly kinetics was reported in our previous paper.[41] However, the scenario at the orthocenter is quite different as we envisage distinct forces involved in this assembly process. Fig. 3b represents the growth kinetics at the orthocenter. It is evident from the plot that the saturation of the nanoparticle assembly had not been achieved within the experimental time limit i.e. ~15 minutes. As the assembly process at the orthocenter depends on the three excitation points and the balance between two counter steering forces, the assembly process is complicated. The effect of the complex nature of this process is perceived from the dynamics of the assembly process, (movie-2 in ESI). It is also important to notice that significant number of nanoparticles escape out from the orthocenter at regular intervals. Furthermore, tracing the central position of the nanoparticle-assembly shows (Fig. 3c) that, the assembly position at the orthocenter is not static, unlike at the excitation spots. The variations in position of the nanoparticles are compared in Fig. 3d and it is apparent that at the orthocenter the nanoparticles assembly fluctuates over a 100μm diameter whereas the assembly at the excitation points is very compact. So this analysis helped us to understand at least the basic mechanism behind the multiple-excitation plasmofluidic assembly process. Though the whole process is byzantine to realize, still one can be curious enough to introduce one more excitation into our experimental configuration.

**Quadruple-spot-excited plasmofluidic assembly of Ag nanoparticles:**

Thus we performed the assembly experiment with four-spot-excitation geometry. Fig. 4a shows the dark-field image of the excitation geometry. In Fig. 4b the FDTD simulation reveals the nearfield distribution due to excitation of four spots. We can see that the SPP interference forms a unique pattern on the plasmonic film, which in turn governs the formation of the nanoparticles chain. Fig. 4c presents the bright-field image of the nanoparticle assembly with the excitation geometry shown in Fig. 4a. For such geometry the assembly process followed the same basic pattern of creating two types of assembly: one at the excitation point and other at the central part. The difference in the nanoparticle density at the four excitation spots occurs due to the slight inequality in power. The extrapolation from three-spots-excitation to four-spots-excitation geometry in plasmofluidic field resulted in interesting patterns of the nanoparticles-assembly and can be further tailored by changing the excitation points. Since this assembly technique is reconfigurable in fluid phase, one can draw any desired sketch of nanoparticles-assembly on a single substrate by merely changing the excitation geometry.

**Aligned assembly of Ag nanowires in plasmofluidic field:**

Next we asked whether this method could be extrapolated to anisotropic nanogeometries. To test the versatility of our method, we implemented this technique to assemble one-dimensional nanostructures. Herein, we used Ag nanowires (prepared by polyol process[43]) suspended in aqueous medium; the assembly process was performed in dual spot excitation geometry and we observed not only the assembly of nanowires but their alignment too. Fig. 5a shows the dark field image of dual spot excitation of the unstructured thin film. The portion marked by dotted box was closely monitored by using × 60-magnification objective lens. The FDTD simulation for similar excitation geometry reveals a typical SPP interference pattern with consecutive maxima and minima, shown in Fig. 5b. This type of SPP interference not only assembles the Ag nanowires, but also helps to align in a particular orientation. Fig. 5c shows the aligned assembly of Ag nanowires on an unstructured film in dual excitation geometry. Thus, we see that the same SPP interference which was attributed to the formation of nanoparticles-chain can be utilized to align much heavier nanowires at the metal-fluid interface.

Next we employed the same experiment with a structured plasmonic film. The gold thin film was patterned as strips of 2 μm width each and the gap between two consecutive strips was 2 μn (see experimental section for more details). In this experiment only a single-spot excitation was used. Fig. 5d is the optical bright and dark-field images of the structured substrate. The nearfield distribution was simulated in FDTD (Fig. 5e) which revealed a pattern of consecutive field maxima and minima which actually aided the nanowire alignment. The bright field image of assembled Ag nanowires is shown in Fig. 5f. The clearly defined field pattern created by the structured thin film assembled and aligned the heavier nanowires with greater precision (movie-3 in ESI). The power density used for this experiment was approximately of 100W/cm$^2$. The two important aspects of this aligned assembly are reversibility and ease. Thus, one can recreate this alignment of nanowires for multiple cycles of experiments with a very low power excitation.

**Conclusion:**

The motivation behind this study was twofold. First, we wanted to explore the possibility of plasmonic-nanoparticle assemblies due to three and four evanescent-wave plasmon excitation at metal-fluid interface. Our results showed that for triple-trap configuration, nanoparticles assembled not only at the excitation points but also at the orthocentre of the triangle formed between the three excitation spots. The reason for such assembly was mainly due to the surface plasmon interference from three different excitations. However, the convective forces resulting from plasmonic heating may also contribute towards assembly formation. Kinetic studies based on particle-tracking at various locations in the assembly indicated a dynamic equilibrium, with the assembly at the orthocentre exhibiting greater fluctuation in its position compared to the excitation spots. For the four-spot configuration, we observed interesting assembly patterns which were mainly governed by the plasmon interference patterns. Numerical simulations for both triple and four spot illuminations confirmed plasmon interference, and were in agreement with our experimental observations. The second motivation of our study was to explore the possibilities of assembling anisotropic plasmonic geometries, such as Ag nanowires, using our method. We found that dual-trap configuration can indeed assemble and weakly align nanowires in liquid, and the directionality of the alignment was governed by interference pattern of counter-propagating surface plasmons at metal-fluid interface. Interestingly, we also found that nanowires can also be assembled and aligned, with better efficiency, using structured metal films and a single evanescent-wave excitation of surface plasmons. All these experiments indicated that plasmon excitation and interference can be harnessed for micron to millimetre scale reversible-assembly of nanoparticles and elongated nanostructures at metal-fluid interface. There are new directions that our work can lead to such as reconfigurable plasmonic metamaterials at metal-fluid interface, where control of various parameters of light - intensity, polarization, phase, etc. - can be controlled in fluids by tailoring the geometrical arrangements of nanoparticle assemblies. Such reconfigurable optical platforms can be further harnessed for nano and micro-optical circuits and devices in fluidic environments. In addition to that, nanoparticle assemblies may be harnessed as test-beds of various plasmofluidic effects where understanding the interaction between surface plasmons and the surrounding fluids is of fundamental interest.

**Acknowledgements**

This research work was partially funded by DTS Nanomission Grant (SR/NM/NS-1141/2012) and and DST Nanoscience Unit Grant (SR/NM/NS-42/2009). G.V.P.K. thanks DST, India, for Ramanujan Fellowship.

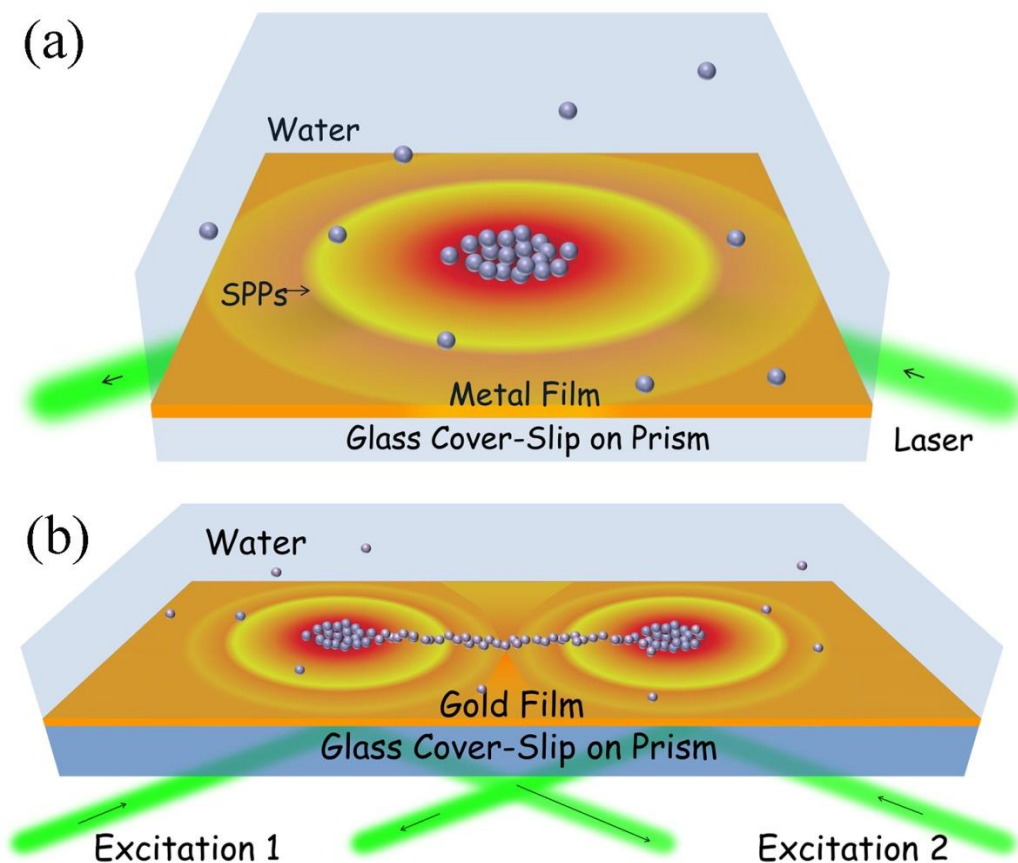

**Fig. 1**: Reprinted from *Nature Communications,* 5, 4357 (2014); (a) Schematic illustration of plasmonic assembly of nanoparticles. Black arrows indicate the direction of 532nm laser (green beam) used for SPP excitation of the metal film deposited over glass coveslip. Grey spheres are the metal nanoparticles dispersed in the water medium. (b) Schematic representation of dual assembly of Ag nanoparticles on Au film coupled to a prism: green lines are the two laser beams of 532nm introduced at two opposite ends of the prism. Black arrows indicate the direction of the laser beams.

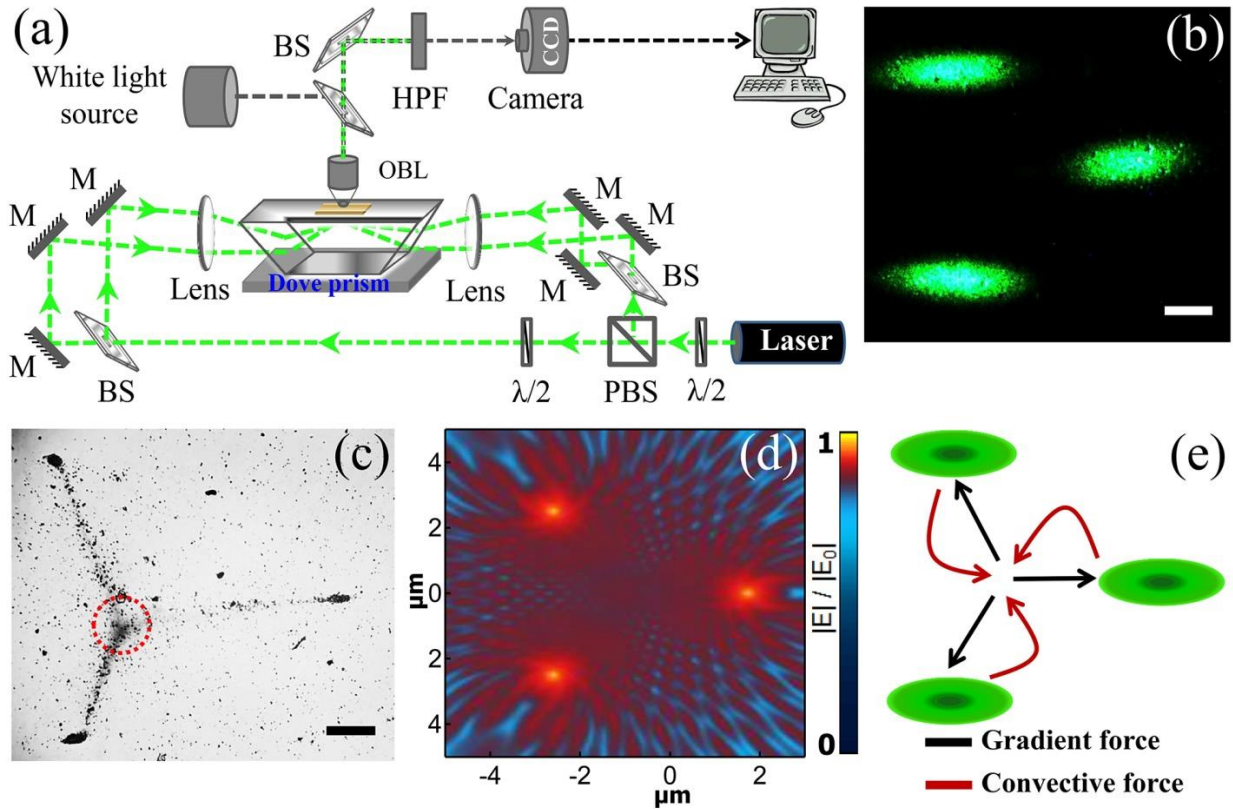

**Fig. 2:** (a) Schematic of the evanescent-wave multi-trap optical set-up used for plasmofluidic assembly of plasmonic nanoparticles. Acronyms: λ/2, half wave plate; PBS, polarizing beam splitter; BS, 50-50 beam splitter; M, mirror; OBL, microscopic objective lens; HPF, high pass filter. (b) Optical dark-field image of the three, evanescently-excited 532nm laser spots on a gold film (50nm thick) optically coupled to the Dove prism. Scale bar is 200 μm. We call this assembly as triple-trap configuration (c) Optical bright-field image of plasmofluidic assembly of Ag nanoparticles at gold film-water interface due to excitation shown in Fig. 2b. The excitation laser was filtered before imaging. Scale bar is 200 μm. Note that the assembly of nanoparticles is denser at the excitation points and at the orthocenter of the three excitation spots indicated by dotted circle. (d) Simulated electric near-field distribution of the three laser spots at gold-water interface. The geometry of the simulation was identical to the experimental configuration. (e) Schematic of the forces involved in the plasmofluidic trapping process.

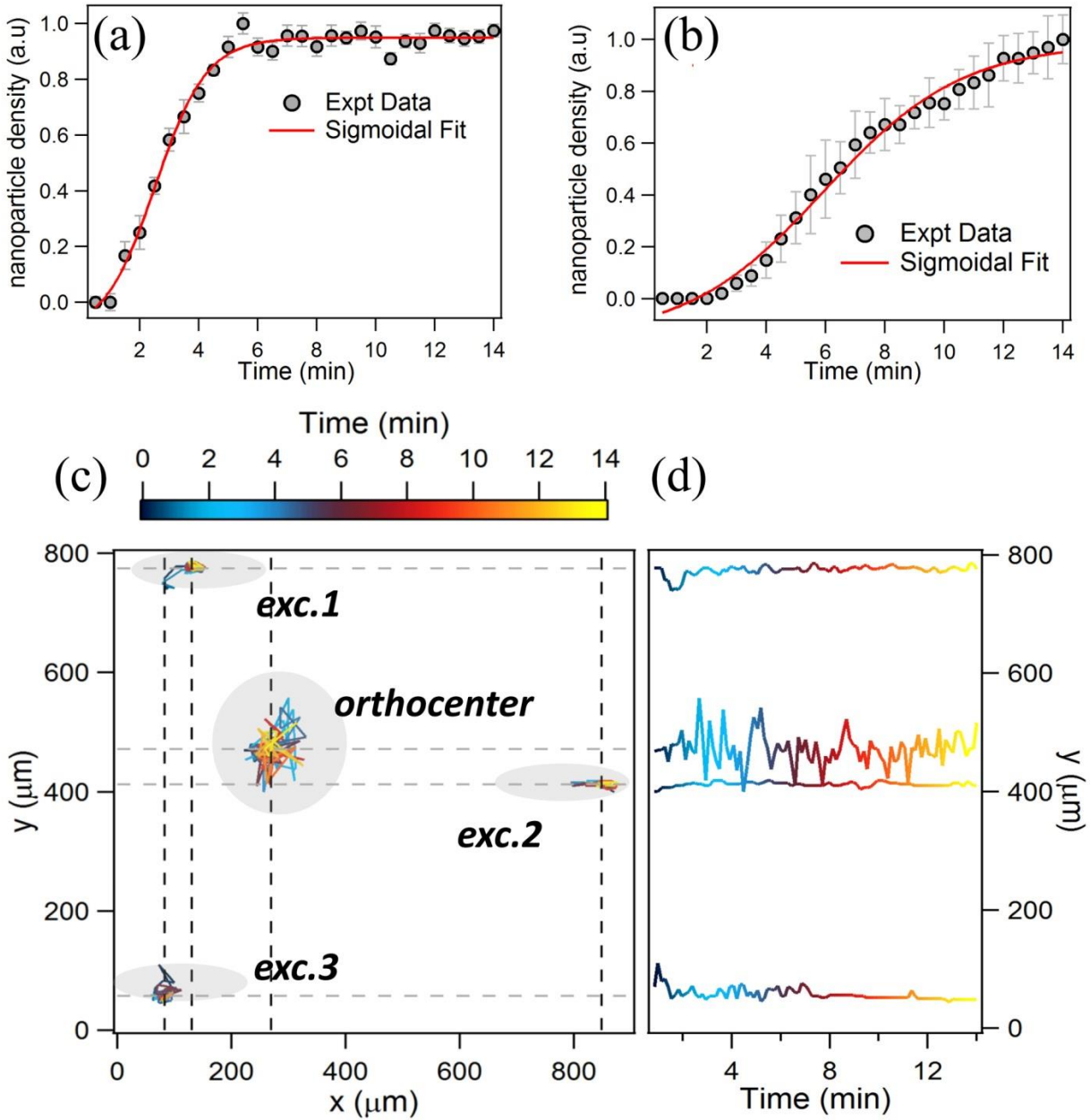

**Fig. 3:** Kinetics of the nanoparticle assembly in the triple-trap configuration. (a) Variation of nanoparticle-density at an excitation point as a function of time. (b) Variation of the nanoparticle density at the orthocenter of the triple-trap. (c) Spatio-temporal variation of the center of nanoparticle assembly at four different locations: excitations 1, 2 and 3, and at the orthocenter. (d) Spatial variation of the same centers along y-axis as function of time. Note that the fluctuation of the center of the orthocenter is more pronounced compared to other locations.

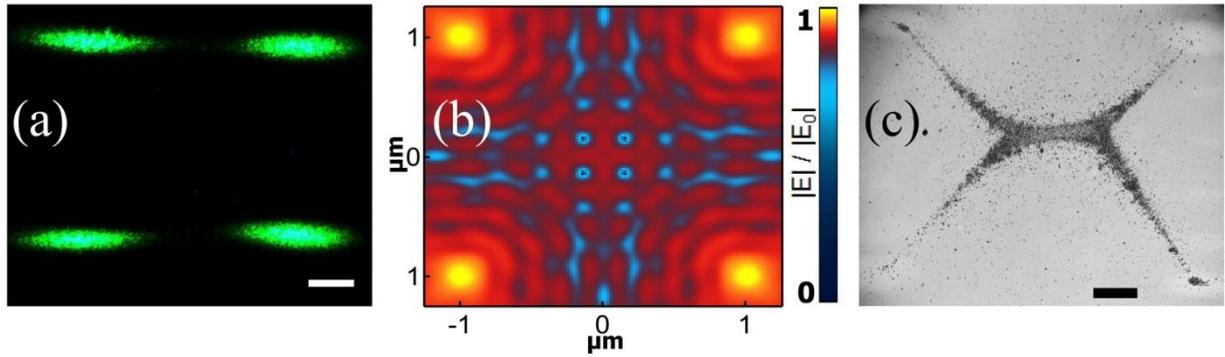

**Fig. 4:** Plasmofluidic assembly of nanoparticles using quadruple-trap configuration. (a) Optical dark-field image of the four, evanescently-excited 532nm laser spots on a gold film (50nm thick) optically coupled to the Dove prism. Scale bar is 200 µm. (b) Simulated electric near-field distribution of the four laser spots at gold-water interface. The geometry of the simulation was identical to the experimental configuration. Plasmonic interference patterns can be clearly observed. (c) Optical bright-field image of plasmofluidic assembly of Ag nanoparticles at gold film-water interface due to excitation shown in Fig. 4a. Scale bar is 200 µm. The nanoparticle densities at the corners of the geometry are different due to small variance in the laser powers.

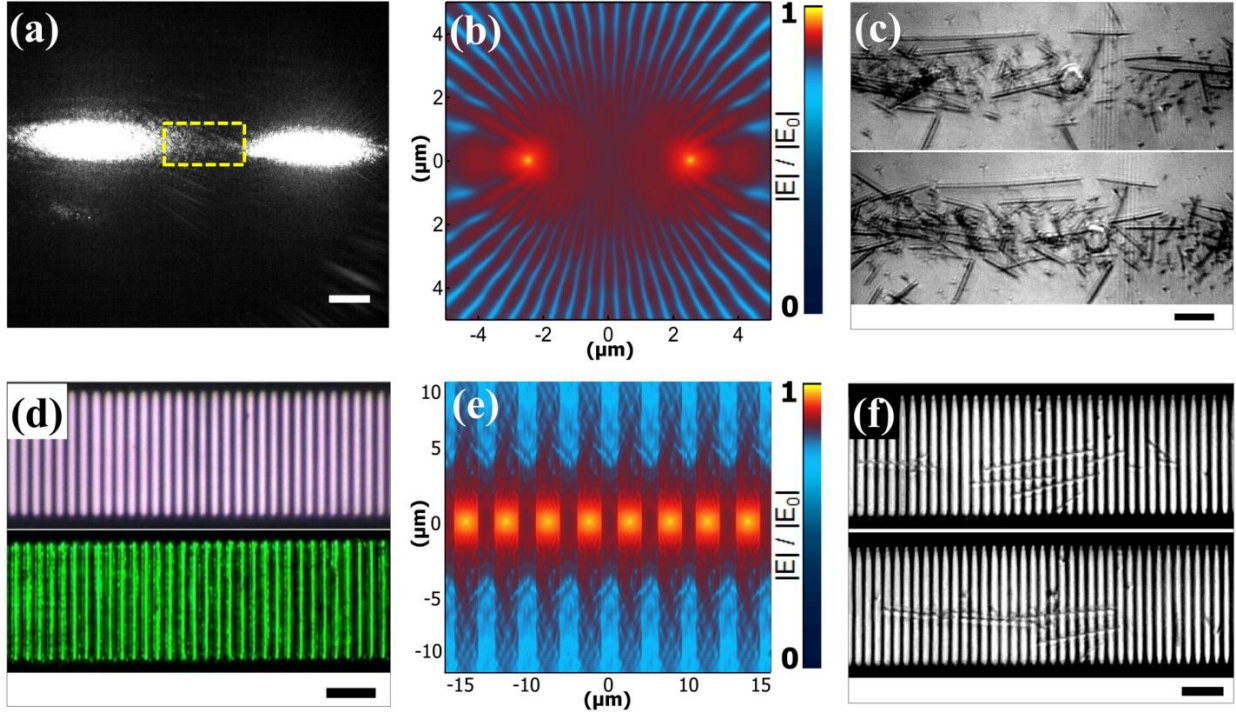

**Fig. 5:** Assembling Ag nanowires using unstructured and structured plasmonic thin film. (a) Optical dark-field image of dual-trap configuration. The region marked with yellow dashed box is the region of interest in our experiments. Scale bar is 200 µm. (b) Simulated electric near-field distribution of the two laser spots at gold-water interface. Plasmonic interference patterns can be clearly observed. (c) Optical bright field images at Ag nanowire assembly at two different instances. We found weak alignment of the longer nanowires. (d) Optical bright-field (top) and dark-field (bottom) images of structured gold film evanescently-excited using a single 532nm laser beam. The length of each Au strip was 50 µm, width was 2 µm and the thickness was 50 nm; and the gap between two consecutive strips was 2 µm. (e) Simulated electric near-field distribution of a single laser spot exciting the gold strip-water interface. (f) Optical bright-field snap-shots of the aligned nanowires at two different instances. We found the trap to be stiffer and confined to smaller area compared to dual-trap configuration. All scale bars are 20 µm long. Note that experiments in first row of the Fig. 5a were performed using dual evanescent-excitation, whereas experiments in second row was performed using a single evanescent excitation.

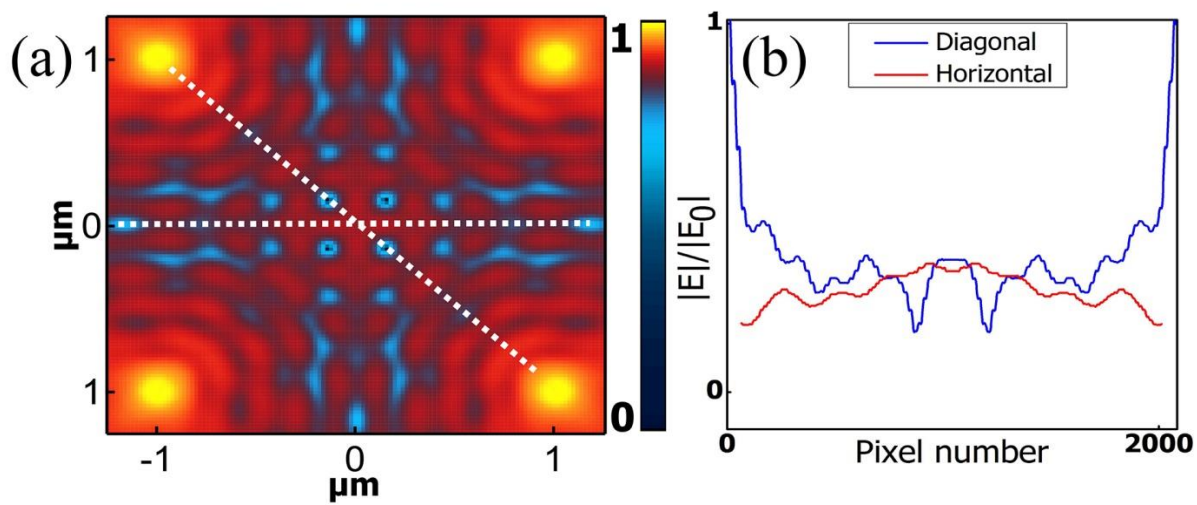

Figure SI-1: The plasmonic field profile along the diagonal and horizontal line (indicated by dotted line) is shown here. It is very clear that the field intensity is more along the diagonal line; thus the nanoparticles assembled along the diagonal in such excitation geometry.

Movie files can be accessed from the below link

http://pubs.rsc.org/en/Content/ArticleLanding/2016/FD/c5fd00127g#!divAbstract